\documentclass[draftcls,onecolumn,letterpaper,11pt]{IEEEtran}

\usepackage{cite,url,color}
\usepackage{graphicx} 
\graphicspath{{./pdf/}{./jpeg/}{./figs/}{./eps/}}
\DeclareGraphicsExtensions{.pdf,.jpeg,.png,.eps}
\usepackage{multirow}
\usepackage[cmex10]{amsmath}
\usepackage{amssymb,bm} 
\usepackage{algorithm,algorithmic}
\usepackage[caption=false,font=footnotesize]{subfig}
\usepackage[amsmath,thmmarks,hyperref]{ntheorem}

\newcommand{\ve}[1]{\mathbf{{#1}}}
\newcommand{\abs}[1]{\ensuremath{\vert #1\vert}}
\newcommand{\norm}[1]{\ensuremath{\Vert #1\Vert}}

\newcommand{\ca}{\ensuremath{\mathrm{Ca}}}
\DeclareMathOperator*{\argmax}{arg\,max}
\DeclareMathOperator*{\argmin}{arg\,min}

\begin{document}

\title{py-oopsi: the python implementation of the fast-oopsi algorithm}
\author{Benyuan~Liu$^{1}$%
\thanks{$^1$Benyuan Liu is with The Department of Biomedical Engineering, Fourth Military Medical University,
Xi'an, 710032, China. E-mail: liubenyuan@gmail.com, lbyoopp@163.com}%
\thanks{Manuscript received \today{}.}}

\markboth{Technical Report, Dept. BME, FMMU, 2014}%
{Liu \MakeLowercase{\textit{et al.}}: The python implementation of fast-oopsi.}



\maketitle

\begin{abstract}
    Fast-oopsi was developed by joshua vogelstein in 2009, which is now widely used to extract neuron spike activities from calcium fluorescence signals. Here, we propose detailed implementation of the fast-oopsi algorithm in python programming language. Some corrections are also made to the original fast-oopsi paper.
\end{abstract}


\ifCLASSOPTIONpeerreview
\begin{center} \bfseries EDICS Category: SAS-MALN \end{center}
\fi
%
\IEEEpeerreviewmaketitle

\section{Fast-oopsi, a brief view}

Oopsi, from vogelstein \cite{vogelstein2010oopsi,vogelstein2010fast}, is a family of optimal optical spike inference algorithms. Here, we focus on the development of the fast-oopsi, which was originally published in \cite{vogelstein2010fast}. We will port the MATLAB implementation to python. Sec \ref{sec:model}, \ref{sec:bayes}, \ref{fast-oopsi:bayes}, \ref{fast-oopsi:solver} and \ref{fast-oopsi:parameters} are digests from the original paper by vogelstein \cite{vogelstein2010oopsi}.

The python implementation, \textcolor{red}{\textbf{py-oopsi}}, can be obtained at \textcolor{red}{\url{https://github.com/liubenyuan/py-oopsi}}.

\section{Calcium fluorescence model}
\label{sec:model}

Let $\ve{F}$ be a one-dimensional fluorescence trace. At time $t$, the fluorescence measurement $F_t$ is a linear Gaussian function of the intracellular calcium concentration $[\ca^{2+}]_t$ at that time:
\begin{equation}
    F_t = \alpha [\ca^{2+}]_t + \beta + e_t, \qquad e_t\sim\mathcal{N}(0,\sigma^2)
\end{equation}
$\alpha$ determines the scale of the signal, $\beta$ absorbs the offset. $\alpha$ and $\beta$ may be learned independently per neuron. The noise $e_t$ is assumed to be i.i.d distributed.

The calcium concentration jumps $A$ $\mu$M after each spike and decays back down to baseline $C_b$ $\mu$M, with time constant $\tau$,
\begin{equation}
    [\ca^{2+}]_{t+1} = (1-\Delta/\tau)[\ca^{2+}]_t + (\Delta/\tau)C_b + An_t,
\end{equation}
where $\Delta$ is the frame interval. The scale $A$ and $\alpha$, baseline $C_b$ and $\beta$ are not identifiable, therefore, we may let $A=1$ and $C_b=0$ without loss of generality. $n_t$ indicates the number of times the neuron spiked in time $t$, we may also write it as a delta function $\delta_t$.

Finally, letting $\gamma=(1-\Delta/\tau)$, we have
\begin{equation}
    C_{t} = \gamma C_{t-1} + n_t
\end{equation}
and (the filtering model)
\begin{equation}
    C[z] = \frac{1}{1-\gamma z^{-1}} N[z]
\end{equation}
Note that $C_t$ does not refer to the absolute intracellular concentration, but rather, a relative measure\cite{vogelstein2010fast}. The simulated calcium trace can be generated if we synthetically generate $n_t$ from a probability distribution. To complete the generative model, we assume spikes are sampled according to a Poisson distribution,
\begin{equation}
    n_t \sim \text{Poisson}(\lambda\Delta)
\end{equation}
where $\lambda\Delta$ is the expected firing rate per bin, $\Delta$ is included to ensure that the expected firing rate is independent of the frame rate\cite{vogelstein2010fast}.

\section{Bayes Model}
\label{sec:bayes}
We aim to find the most likely spike trains $\hat{\ve{n}}$ given the fluorescence $\ve{F}$,
\begin{equation}\label{eq:n_map0}
    \hat{\ve{n}} = \argmax_{n_t\in\mathcal{N}_0,\forall t} p(\ve{n}|\ve{F})
\end{equation}
Using Bayes' rule,
\begin{equation}
    p(\ve{n}|\ve{F}) = \frac{1}{p(\ve{F})}\cdot p(\ve{F}|\ve{n})p(\ve{n})
\end{equation}
given that $p(\ve{F})$ merely scales the results, we rewrite \eqref{eq:n_map0} as,
\begin{equation}
    \hat{\ve{n}} = \argmax_{n_t\in\mathcal{N}_0,\forall t} p(\ve{F}|\ve{n})p(\ve{n})
\end{equation}
and we already have,
\begin{align}
    p(\ve{F}|\ve{n}) &= p(\ve{F}|\ve{C}) = \prod p(F_t|C_t), \\
    p(\ve{n}) &= \prod p(n_t),
\end{align}
where,
\begin{align}
    p(F_t|C_t) &= \mathcal{N}(\alpha C_t + \beta,\sigma^2), \\
    p(n_t) &= \mathrm{Poisson}(\lambda\Delta)
\end{align}
The Poisson distribution penalize sparsity (a sparse prior).

Finally, we have the cost function,
\begin{align}
    \hat{\ve{n}} &= \argmax_{n_t\in\mathcal{N}_0} \prod_{t=1}^{T} \frac{1}{\sqrt{2\pi\sigma^2}}\exp\left\{ -\frac{1}{2}\frac{(F_t - \alpha C_t - \beta)^2}{\sigma^2}\right\} \frac{\exp\left\{ -\lambda\Delta \right\}(\lambda\Delta)^{n_t}}{n_t!} \\
                 &= \argmax_{n_t\in\mathcal{N}_0} \sum_{t=1}^{T} -\frac{1}{2\sigma^2}(F_t - \alpha C_t - \beta)^2 + n_t\ln \lambda\Delta - \ln n_t!
\end{align}
However, solving for this discretized optimization problem is computational intractable.

\section{Approximate Bayes Filter}
\label{fast-oopsi:bayes}

We can approximate the Poisson distribution with an exponential distribution of the same mean,
\begin{equation}
    \frac{\exp\{-\lambda\Delta\}(\lambda\Delta)^{n_t}}{n_t!} \rightarrow (\lambda\Delta)\exp\{-n_t\lambda\Delta\}
\end{equation}
and consequently,
\begin{equation}
    \hat{\ve{n}} = \argmax_{n_t>0} \sum_{t=1}^{T} -\frac{1}{2\sigma^2}(F_t - \alpha C_t - \beta)^2 - n_t\lambda\Delta
\end{equation}
note that $n_t\in\mathcal{N}_0$ has been replaced by $n_t>0$, since exponential distribution can yield any nonnegative number\cite{vogelstein2010fast}. The exponential approximation imposes a sparsening effect, and also, it makes the optimization problem concave in $\ve{C}$, meaning that any gradient descent algorithm guarantees achieving \textbf{the global maxima} (because there are no local minima).

We may further drop the constraint (nonnegative) by adopting \textbf{interior point} method,
\begin{equation}
    \hat{\ve{n}} = \argmax_{n_t} \sum_{t=1}^{T} -\frac{1}{2\sigma^2}(F_t - \alpha C_t - \beta)^2 - n_t\lambda\Delta + z\ln n_t
\end{equation}
where we add a weighted barrier term that approaches $-\infty$ as $n_t$ approaches zero, by solving for a series of $z$ going down to nearly zero. The goal is to efficiently solve,
\begin{equation}
    \hat{\ve{C}} = \argmax_{C} \sum_{t=1}^{T} -\frac{1}{2\sigma^2}(F_t - \alpha C_t - \beta)^2 - (C_t - \gamma C_{t-1})\lambda\Delta + z\ln (C_t - \gamma C_{t-1})
\end{equation}
this cost function is twice differentiable, one can use the Newton-Raphson technique to ascend the surface.

\section{Matrix Notation and the Newton-Raphson solver}
\label{fast-oopsi:solver}
To proceed, we have
\begin{equation}
    \ve{M}\ve{C} =
    \begin{bmatrix}
        -\lambda & 1 & 0 & 0 & \cdots & 0 \\
        0 & -\lambda & 1 & 0 & \cdots & 0 \\
        \vdots & \ddots & \ddots & \ddots & \ddots & \vdots \\
        0 & \cdots & 0 & -\lambda & 1 & 0 \\
        0 & \cdots & 0 & 0 & -\lambda & 1
    \end{bmatrix}
    \begin{bmatrix}
        C_1 \\
        C_2 \\
        \vdots \\
        C_{T-1} \\
        C_{T}
    \end{bmatrix} =
    \begin{bmatrix}
        n_1 \\
        n_2 \\
        \vdots \\
        n_{T-1}
    \end{bmatrix}
\end{equation}
$\ve{M}$ is a $(T-1)\times T$ matrix. Now letting $\ve{1}$ be a $(T-1)\times 1$ column vector, $\bm{\lambda} = (\lambda\Delta)\ve{1}$, $\bm{\alpha}$ and $\bm{\beta}$ a $T$-dimensional vector, $\odot$ to indicate element-wise operations, then \footnote{contrary to \cite{vogelstein2010fast}, but alike \textbf{fast-oopsi.m}, we choose $\ve{M}$ as a sparse $T\times T$ matrix, and $\ve{1}$ as $T\times 1$ column vector. Therefore we have $n_0=C_0$, we will correct $n_0=\epsilon$ after convergence.}
\begin{equation}
    \hat{\ve{C}} = \argmax_{\ve{MC}\geq_{\odot}\ve{0}} -\frac{1}{2\sigma^2}\norm{\ve{F}-\bm{\alpha}\ve{C}-\bm{\beta}}_2^2 - (\ve{MC})^T\bm{\lambda} + z\ln_{\odot} (\ve{MC})^T\ve{1}
\end{equation}

We instead iteratively minimize the cost function $\mathcal{L}$ (called \textit{post} in our python implementation) where,
\begin{equation}
    \hat{\ve{C}}_z = \argmin_{\ve{C}} \mathcal{L}, \quad \mathcal{L}= \frac{1}{2\sigma^2}\norm{\ve{F}-\bm{\alpha}\ve{C}-\bm{\beta}}_2^2 + (\ve{MC})^T\bm{\lambda} - z\ln_{\odot} (\ve{MC})^T\ve{1}
\end{equation}
$\mathcal{L}$ is \textbf{convex}, when using Newton-Raphson method to \textbf{descend} a surface, one iteratively computes the gradient $\ve{g}=\nabla\mathcal{L}$ (first derivative) and Hessian $\ve{H}=\nabla^2\mathcal{L}$ (second derivative) of the argument to be optimized. Then, $\ve{C} = \ve{C} - s\ve{d}$, where $s$ is the step size and $\ve{d}$ is the step direction by solving $\ve{H}\ve{d} = \ve{g}$. The gradient and Hessian, with respect to $\ve{C}$, are
\begin{align}
    \ve{g} &= -\frac{\bm{\alpha}}{\sigma^2} (\ve{F} - \bm{\alpha}\ve{C} - \bm{\beta}) + \ve{M}^T\bm{\lambda} - z\ve{M}^T(\ve{MC})_{\odot}^{-1} \\
    \ve{H} &= \frac{\alpha^2}{\sigma^2}\ve{I} + z\ve{M}^T(\ve{MC})_{\odot}^{-2}\ve{M}
\end{align}
$s$ is found via \textbf{backtracking linesearches}. $\ve{M}$ is bidiagonal, so $\ve{H}$ is tridiagonal, $\ve{d} = \ve{H}^{-1}\ve{g}$ can be efficiently implemented in matlab by assuming $\ve{H}$ is a sparse matrix. In python, we may use sparse linsolvers (linsolve.spsolve) to efficiently find $\ve{d}$. Once $\hat{\ve{C}}$ is obtained, it is a simple linear transform to obtain $\hat{\ve{n}}$, via $\hat{\ve{n}}=\ve{M}\hat{\ve{C}}$. We will normalize $\ve{n}$ by $\ve{n}=\ve{n}/\mathrm{max}(\ve{n})$ after convergence.

\section{Parameters initialize and update}
\label{fast-oopsi:parameters}

The parameters $\bm{\theta}=\{\alpha,\beta,\sigma,\gamma,\lambda\}$ are unknown. We may use pseudo expectation-maximization method,
(1), initialize the parameters, (2) recursively computes $\hat{\ve{n}}$ and updating $\bm{\theta}$ given the new $\hat{\ve{n}}$ until the convergence is met.

The scale of $\ve{F}$ relative to $\ve{n}$ is arbitrary, therefore, $\ve{F}$ is firstly \textit{detrended}, and then \textit{linearly mapped} between $0$ and $1$.
\begin{equation}
    \ve{F} = \mathrm{detrend}(\ve{F}), \quad \ve{F}=(\ve{F} - F_{min}) / (F_{max} - F_{min}),
\end{equation}
Next, because spiking is sparse in many experimental settings, $\ve{F}$ tends to be around baseline, $\beta$ is set to the median of $\ve{F}$. We use median absolute deviation (MAD) and correction factor $K$, as a robust normal scale estimator of $\ve{F}$ where $K=1.4826$. Previous works showed that the results $\hat{\ve{n}}$ and $\hat{\ve{C}}_z$ are robust to minor variations in the time constant, we let $\gamma=1-\Delta$. Finally, $\lambda$ is set to $1$Hz, which is between baseline and evoked spike rate for data of interest \footnote{corrections to \cite{vogelstein2010fast}: 1), add detrend to $\ve{F}$, 2), $K=1.4826$ and it is multiplied (not divided by) $\mathrm{MAD}(\ve{F})$.}.
\begin{align}
    \alpha &= 1, \\
    \beta &= \mathrm{median}(\ve{F}), \\
    \sigma &= \mathrm{MAD}(\ve{F})\cdot K = \mathrm{median}(\abs{\ve{F} - \beta})\cdot K, \quad K = 1.4826 \\
    \gamma &= 1 - \Delta/(1 \mathrm{sec}), \\
    \lambda &= 1 \mathrm{Hz}
\end{align}

Then, given $\hat{\ve{C}}$ and $\hat{\ve{n}}$, we may (approximately) update $\bm{\theta}$ by,
\begin{equation}
    \hat{\bm{\theta}} \approx \argmax_{\bm{\theta}} p(\ve{F},\hat{\ve{C}}|\bm{\theta}) = \argmax_{\bm{\theta}} \ln p(\ve{F}|\hat{\ve{C}};\{\alpha,\beta,\sigma\}) + \ln p(\hat{\ve{n}}|\lambda)
\end{equation}
where,
\begin{align}
    \hat{\lambda} &= \argmax_{\lambda>0} \sum_{i=1}^T \left[ \ln(\lambda\Delta) + \hat{n_t}\lambda\Delta \right] \\
    \{\hat{\alpha},\hat{\beta},\hat{\sigma}\} &= \argmax_{\alpha,\beta,\sigma >0}\sum_{i=1}^T \left[ -\frac{1}{2}\ln(2\pi\sigma^2) - \frac{1}{2}\left( \frac{F_t - \alpha C_t - \beta}{\sigma} \right)^2 \right]
\end{align}
We have (by taking the derivatives and letting them equal zero),
\begin{align}
    \hat{\lambda} &= \frac{T}{\Delta \sum_t n_t}, \\
    \hat{\alpha} &= 1, \\
    \hat{\beta} &= \frac{\sum_t (F_t - C_t)}{T}, \\
    \hat{\sigma}^2 &= \frac{\sum_t (F_t - C_t - \beta)^2}{T} = \frac{\norm{\ve{F} - \ve{C} - \bm{\beta}}_2^2}{T}
\end{align}
where $\hat{\lambda}$ is the inverse of the inferred firing rate, $\hat{\alpha}$ can be set to $1.0$ because the scale of $\ve{C}$ is arbitrary, $\hat{\beta}$ is the mean bias, $\hat{\sigma}$ is the root-mean-square of the residual error.

\section{Implementation of oopsi}
Matlab implementation is available, here we focus on the python migrant, and correct some typos in \cite{vogelstein2010fast} as needed. The python code itself explains all, see \ref{fast-oopsi:bayes}, \ref{fast-oopsi:solver}, and \ref{fast-oopsi:parameters} for detailed documentary. Pseudo code can be found in Algo \ref{algo:fast-oopsi}. Algo \ref{algo:oopsi_est_map} describe the subroutine \textbf{MAP}, Algo \ref{algo:oopsi_est_par} describe the subroutine \textbf{update}.
\begin{algorithm}[htbp]
    \caption{Pseudo code (python) for fast-oopsi}\label{algo:fast-oopsi}
    \begin{algorithmic}[1]
        \STATE Initialize parameters $\ve{P}$: $\ve{F}=\mathrm{detrend}(\ve{F})$, $\ve{F}=(\ve{F}-\mathrm{min}(\ve{F}))/(\mathrm{max}(\ve{F})-\mathrm{min}(\ve{F}))$, $\alpha=1.0$, $\beta=\mathrm{median}(\ve{F})$, $\lambda=1.0$, $\gamma=1-\Delta$, $\sigma=\mathrm{MAD}(\ve{F})\cdot 1.4826$, $T=\mathrm{len}(\ve{F})$
        \STATE one-shot Newton-Raphson $\ve{n}$, $\ve{C}$, $\mathcal{L}$ = MAP($\ve{F}$,$\ve{P}$), see Algo \ref{algo:oopsi_est_map}.
        \FOR{$i$ in $1\cdots\mathrm{iterMax}$}
        \STATE update parameters $\ve{P}$ = update($\ve{n}$,$\ve{C}$,$\ve{F}$,$\ve{P}$), see Algo \ref{algo:oopsi_est_par}.
        \STATE iterative through $\ve{n}$, $\ve{C}$, $\mathcal{L}$ = MAP($\ve{F}$,$\ve{P}$),
        \STATE let $\mathcal{L}^{(k)}=\{\mathcal{L}_1,\cdots,\mathcal{L}_k\}$
        \IF{$\abs{\frac{\mathcal{L}_i-\mathcal{L}_{i-1}}{\mathcal{L}_i}}<\mathrm{ltol}$ or $\mathrm{any}(\abs{\mathcal{L}^{(i)}-\mathcal{L}_i})<\mathrm{gtol}$}
        \STATE break
        \ENDIF
        \ENDFOR
    \end{algorithmic}
\end{algorithm}

\begin{algorithm}[htbp]
    \caption{Pseudo code of subroutine \textit{MAP}}\label{algo:oopsi_est_map}
    \begin{algorithmic}[1]
        \STATE Initialize $\ve{n}=0.01\ve{1}$
        \STATE Initialize $\ve{C}(z) = 1/(1-\gamma) \ve{N}(z)$
        \STATE Initialize $\bm{\lambda}=\lambda\Delta \ve{1}$
        \FOR{$z=1.0$, $z>1e-13$, $z=z/10$}
            \STATE calculate $\mathcal{L}_z$
            \WHILE{$s>1e-3$ or $\norm{\ve{d}}>5e-2$}
            \STATE Calculate $\ve{g}$, $\ve{H}$ and $\ve{d}=\mathrm{spsolve}(\ve{H},\ve{g})$
            \STATE Find $s$ : $\ve{h}=-\ve{n}/(\ve{M}\ve{d})$, $s=\min (0.99\ve{s}[\ve{s}>0],1.0)$
            \STATE Initialize $\mathcal{L}_s=\mathcal{L}_z + 1$
                \WHILE{$\mathcal{L}_s>\mathcal{L}_z + 1e-7$}
                \STATE $\ve{C}=\ve{C}+s\ve{d}$
                \STATE $\ve{n}=\ve{MC}$
                \STATE update $\mathcal{L}_s$
                \STATE decrease $s=s/5.0$
                \IF{$s<1e-20$}
                \STATE break
                \ENDIF
                \ENDWHILE
            \ENDWHILE
        \ENDFOR
    \end{algorithmic}
\end{algorithm}

\begin{algorithm}[htbp]
    \caption{Pseudo code of subroutine \textit{update}}\label{algo:oopsi_est_par}
    \begin{algorithmic}[1]
        \STATE $\alpha=1.0$
        \STATE $\beta=\sum_i (F_i-C_i)/T$
        \STATE $\sigma^2=\norm{\ve{F}-\alpha\ve{C}-\beta}_2^2/T$
        \STATE $\lambda=T / (\Delta\sum_i n_i)$
    \end{algorithmic}
\end{algorithm}

\section{Wiener filter (linear regression, simple convex optimization)}
In the wiener filter, we approximate the Poisson distribution with a Gaussian distribution,
\begin{equation}
    p(n_t) \sim \mathcal{N}(\lambda\Delta,\lambda\Delta)
\end{equation}
then, the MAP estimator yields,
\begin{equation}
    \hat{\ve{n}} = \argmax_{n_t} \sum_{t=1}^T \left[ -\frac{1}{2\sigma^2}(F_t - \alpha C_t -\beta)^2 -\frac{1}{2\lambda\Delta}(n_t - \lambda\Delta)^2 \right]
\end{equation}
and its matrix notation,
\begin{equation}
    \hat{\ve{C}} = \argmax_{\ve{C}} -\frac{1}{2\sigma^2}\norm{\ve{F} - \alpha\ve{C} - \beta\ve{1}}_2^2 - \frac{1}{2\lambda\Delta}\norm{\ve{MC}-\lambda\Delta\ve{1}}_2^2
\end{equation}
which is \textbf{quadratic}, \textbf{concave} in $\ve{C}$.

Finally, we aim to optimize (minimize, \textbf{quadratic}, \textbf{convex} optimization),
\begin{equation}
    \hat{\ve{C}} = \argmin_{\ve{C}} \mathcal{L}, \quad \mathcal{L}= \frac{1}{2\sigma^2}\norm{\ve{F} - \alpha\ve{C} - \beta\ve{1}}_2^2 + \frac{1}{2\lambda\Delta}\norm{\ve{MC}-\lambda\Delta\ve{1}}_2^2
\end{equation}
where $\mathcal{L}$ is \textbf{convex} in $\ve{C}$. Using Newton-Raphson update, we find $\ve{C}=\ve{C}-\ve{d}$, $\ve{H}\ve{d}=\ve{g}$ and $\ve{g}=\nabla \mathcal{L}$, $\ve{H}=\nabla^2\mathcal{L}$. The gradient $\ve{g}$ and Hessian $\ve{H}$ are,
\begin{align}
    \ve{g} &= -\frac{\alpha}{\sigma^2}(\ve{F}-\alpha\ve{C}-\beta\ve{1}) + \frac{1}{\lambda\Delta}\left[ \ve{M}^T(\ve{MC}) + \lambda\Delta\ve{M}^T\ve{1}\right] \\
    \ve{H} &= \frac{\alpha^2}{\sigma^2}\ve{I} + \frac{1}{\lambda\Delta}\ve{M}^T\ve{M}
\end{align}
In the python implementation, we let $\alpha=1.0$ and $\beta=0.0$. Pseudo code can be found in Algo \ref{algo:fast-wiener}.
\begin{algorithm}[htbp]
    \caption{Pseudo code (python) for wiener filter}\label{algo:fast-wiener}
    \begin{algorithmic}[1]
        \STATE Initialize $\ve{F}=(\ve{F}-\mathrm{mean}(\ve{F}))/\mathrm{max}(\abs{\ve{F}})$, $\sigma=0.1\norm{\ve{F}}_2$
        \STATE Calculate $\mathcal{L}_0$
        \FOR{$i$ in $1\cdots$ iterMax}
        \STATE Calculate $\ve{g}$, $\ve{H}$ and $\ve{d}=\mathrm{spsolve}(\ve{H},\ve{g})$
        \STATE Calculate $\ve{C}=\ve{C} - \ve{d}$
        \STATE Calculate $\mathcal{L}_i$
        \IF{$\mathcal{L}_{i}<\mathcal{L}_{i-1}+\mathrm{gtol}$}
        \STATE $\ve{n}=\ve{N}$
        \STATE $\sigma=\sqrt{\norm{\ve{F}-\ve{C}}_2^2/T}$
        \ENDIF
        \ENDFOR
        \STATE $\ve{n}=\ve{n}/\mathrm{max}(\ve{n})$
    \end{algorithmic}
\end{algorithm}

\section{Simulation Results}
\begin{figure}[htbp]
    \centering
    \includegraphics[width=5in]{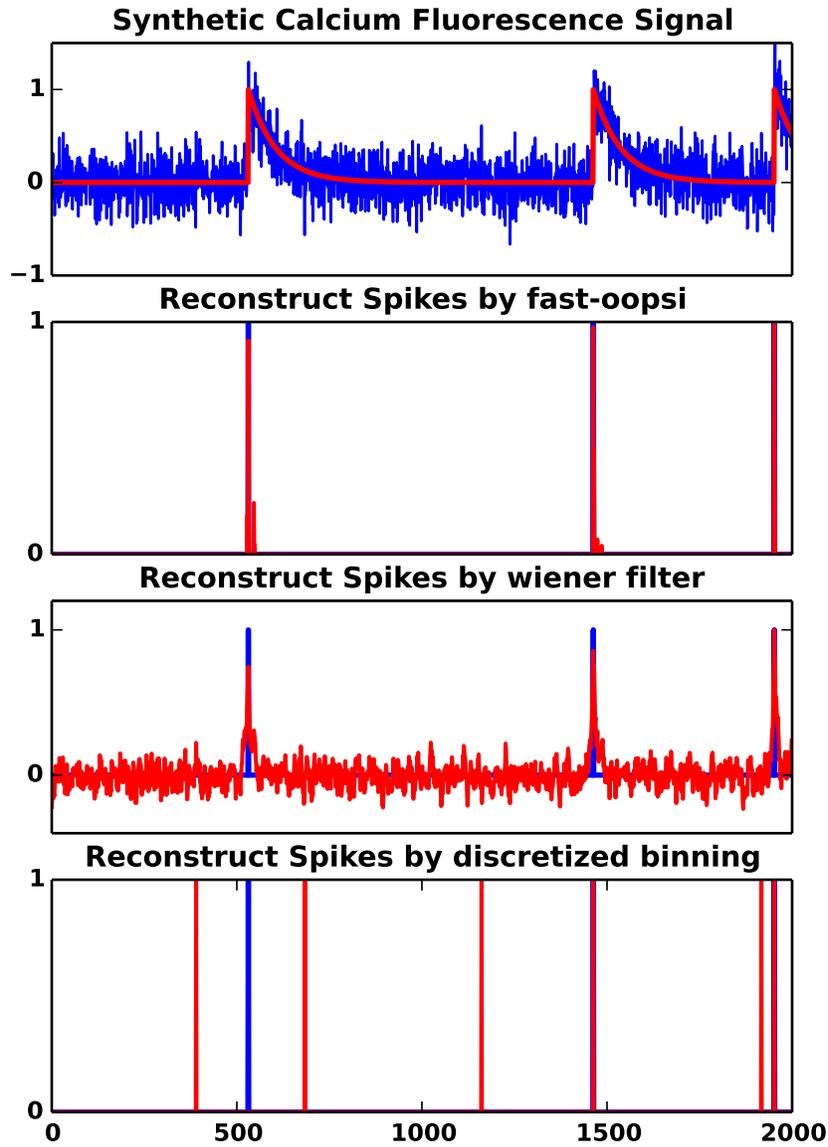}
    \caption{Reconstruct spikes from calcium fluorescence. (a) The synthetic calcium trace. (b), (c), (d) are reconstructed spikes by py-oopsi, wiener filter and discretized binning, respectively.}
    \label{fig:demo}
\end{figure}

We generated synthetic calcium traces with $T=2000$, $\Delta=20$ms, $\lambda=0.1$, $\tau=1.5$. Randomized noise were added with $0.2$ standard deviation. Py-oopsi and wiener filter are used to reconstruct the spikes from calcium fluorescence, where only $\Delta$ is known a prior. The results are shown in Figure \ref{fig:demo}

\ifCLASSOPTIONcaptionsoff
  \newpage
\fi

\bibliographystyle{IEEEtran}
\bibliography{connectomics}
%



%






\end{document}